\RequirePackage{ifpdf}
\documentclass{PoS}
\usepackage[authoryear,square]{natbib}
\bibpunct{(}{)}{;}{a}{}{,}

\title{Searching for Extraterrestrial Intelligence with the Square Kilometre Array}

\ShortTitle{SETI with the SKA}

\author{
\speaker{Andrew P. V. Siemion}$^{1,2,3}$, 
James Benford$^{4}$,
Jin Cheng-Jin$^{5}$,
Jayanth Chennamangalam$^{6}$,
James Cordes$^{7}$,
David R. DeBoer$^{3}$,
Heino Falcke$^{2,1,8,9}$,
Mike Garrett$^{1,10}$,
Simon Garrington$^{11}$,
Leonid Gurvits$^{12,13}$,
Melvin Hoare$^{14}$,
Eric J. Korpela$^{3}$,
Joseph Lazio$^{15}$,
David Messerschmitt$^{3}$,
Ian S. Morrison$^{16}$,
Tim O'Brien$^{10}$,
Zsolt Paragi$^{12}$,
Alan Penny$^{17}$,
Laura Spitler$^{7}$,
Jill Tarter$^{18}$,
Dan Werthimer$^{3}$
\\
$^1$ASTRON, NL;
$^2$Radboud University, NL; 
$^3$University of California, Berkeley, US;
$^4$Microwave Sciences, US;
$^5$NAOC, CN;
$^6$Oxford University, UK;
$^7$Cornell University, US;
$^8$MPIfR, DE;
$^9$NIKHEF, NL;
$^{10}$Leiden University, NL;
$^{11}$Jodrell Bank Observatory, UK;
$^{12}$JIVE, NL;
$^{13}$Delft University of Technology, NL
$^{14}$University of Leeds, UK;
$^{15}$Jet Propulsion Laboratory, California Inst. of Technology, US;
$^{16}$University of New South Wales, AU;
$^{17}$University of St. Andrews, UK;
$^{18}$SETI Inst., US
\\
E-mail: \email{siemion at astron.nl}
}

\abstract{
The vast collecting area of the Square Kilometre Array (SKA), harnessed by sensitive receivers, flexible digital electronics and increased computational capacity, could permit the most sensitive and exhaustive search for technologically-produced radio emission from advanced extraterrestrial intelligence (SETI) ever performed.  For example, SKA1-MID will be capable of detecting a source roughly analogous to terrestrial high-power radars (e.g. air route surveillance or ballistic missile warning radars, EIRP\footnote{EIRP = equivalent isotropic radiated power} $\sim10^{17} \rm{\ erg\ sec}^{-1}$) at 10 pc in less than 15 minutes, and with a modest four beam SETI observing system could, in one minute, search every star in the primary beam out to $\sim$100 pc for radio emission comparable to that emitted by the Arecibo Planetary Radar (EIRP $\sim2\rm{\ x\ }10^{20} \rm{\ erg\ sec}^{-1}$)\footnote{Both for Band 1}.  The flexibility of the signal detection systems used for SETI searches with the SKA will allow new algorithms to be employed that will provide sensitivity to a much wider variety of signal types than previously searched for.

\vspace{3px}
Here we discuss the astrobiological and astrophysical motivations for radio SETI and describe how the technical capabilities of the SKA will explore the radio SETI parameter space.  We detail several conceivable SETI experimental programs on all components of SKA1, including commensal, primary-user, targeted and survey programs and project the enhancements to them possible with SKA2.  We also discuss target selection criteria for these programs, and in the case of commensal observing, how the varied use cases of other primary observers can be used to full advantage for SETI.\\ 
}

\FullConference{
Advancing Astrophysics with the Square Kilometre Array\\
June 8-13, 2014\\
Giardini Naxos, Sicily, Italy}

\newcommand{\skipthis}[1]{}

\begin{document}

\section{Background}
\subsection{Signatures of Life}
Whether or not life occurs elsewhere in the universe is one of the most profound questions science can ask.  As of this moment, we know of only a single example of the emergence of life, despite evidence that the necessary abiotic precursors and environmental conditions are common.  Earth-like planets \citep{2013PNAS..11019273P}, water \citep{2011Sci...334..338H} and complex chemistry \citep{2009ARA&A..47..427H} have now been found in abundance, emboldening the field of astrobiology and underscoring the allure of extraterrestrial life.  Knowing that extraterrestrial life {\it could} exist, the race is on to discover whether or not it, in fact, {\it does exist}.  Yet more compelling is the possibility that extraterrestrial evolution may have followed a similar track as on the Earth, and brought forth a life form possessing intelligence and a technological capability perhaps far exceeding our own.  

Although spectroscopy of extrasolar planet atmospheres or in-situ sampling missions in our Solar System may soon yield indirect evidence for life beyond Earth, the deeper question of the parameters of life's evolution that might lead to intelligence is addressable only by much more select means.  The creation of technology, and specifically environmental modification by that technology, is the only known tracer of intelligence detectable over interstellar distances.  However, while many chemical tracers of basic life can be ambiguous \citep{2004Icar..172..537K, 2014PNAS..111.6871R}, there exist types of electromagnetic emissions that could plausibly be generated by advanced technology but are not known to ever arise naturally.  As far as we know, these emissions are definite and conclusive indicators of advanced technology, and presumably its intelligent creator.  Radio communication in particular is a superb example of such a probe of extraterrestrial technology and is in fact the most detectable distant signature of our own technology.     

Radio astronomy has long played a prominent role in searches for extraterrestrial intelligence (SETI), beginning with the first suggestions by \cite{1959Natur.184..844C} that narrow-band radio signals near 1420 MHz might be effective tracers of advanced technology and early experiments along these lines by \cite{Drake:1961bv}, continuing through to more recent investigations searching for a variety of coherent radio signals indicative of technology at a wider range of frequencies, e.g. \cite{Tarter:1980p1516, 1986Icar...67..525H, Horowitz:1993p1523, 1994Icar..107..215S, Leigh:1998p12013, Korpela:2002p3225, Gray:2008ws, Rampadarath:2012fw, Siemion:2010p6845, 2013ApJ...767...94S}.  The motivation for radio searches for extraterrestrial intelligence has been throughly discussed in the literature \citep{NASA:2003p185, Tarter:2003p266}, but the salient arguments are the following: 1.  coherent radio emission is commonly produced by advanced technology (judging by Earth's technological development), 2. electromagnetic radiation can convey information at the maximum velocity currently known to be possible, 3. radio photons are energetically cheap to produce, 4. certain types of coherent radio emissions are easily distinguished from astrophysical background sources, 5. these emissions can transit vast regions of interstellar space relatively unaffected by gas, plasma and dust.  These arguments are unaffected by varying assumptions about the motivation of the transmitting intelligence, e.g. whether the signal transmitted is intentional or unintentional, and can be applied roughly equally to a variety of potential signal types or modulation schemes.

\subsection{Artificial Radio Emission}
\label{sec:are}
The first modern radio SETI experiments, including Frank Drake's pioneering work described in \cite{Drake:1961bv}, focused on identifying continuous narrow-band emission.  In the SETI context, ``narrow-band'' emission is defined as electromagnetic emission that is spectrally narrower than natural astrophysical sources.  In the radio regime, the spectrally narrowest natural sources are astrophysical masers, with a spectral width no narrower than $\sim$ 500 Hz.  The vast majority of radio SETI searches conducted since have continued this paradigm.  The motivations for searching for signals of this type are numerous, but prominent among them are that narrow-band signals are readily distinguishable from natural sources, are relatively simple and are easy to detect with current technology.  Were an advanced civilization interested in intentionally signaling another species, these properties would be very valuable.  Narrow-band sinusoids are also a prominent component of many of our terrestrial radio communication and ranging systems, e.g. carrier tones and continuous-wave (CW) radar, but this is changing as terrestrial technology becomes more sophisticated. 

More recently, it has been suggested that various other types of signals possess merit as well.  Several authors have presented variations on the idea that a plausible signal from a very advanced civilization might appear to be a subtle variation on natural emission, either to make use of a natural source's inherent luminosity or to attract the attention of other astronomers \citep{1993ASPC...47..257C, 1994Ap&SS.214..209L, 1995ASPC...74..325C, Sullivan:1995wq, 2015NewA...34..245C}.  \cite{Benford:2010p615,2010AsBio..10..491B,2010AsBio..10..475B} have argued that the terrestrial economics of beacon transmitter construction and operation imply that broadband rather than narrow-band systems might be preferred.  \cite{2011AcAau..69..777F} concluded that broadband emission employing frequency-shift keying (FSK) would be preferred by energy-conscious civilizations seeking to transmit information-bearing signals using similar capabilities to our own.  \cite{Messerschmitt:2011ir, 2012AcAau..78...80M} similarly arrived at the idea that broadband communication might be preferable by considering robustness to radio frequency interference (RFI), and further deduced that this emission may have time-bandwidth extents influenced by the properties of the interstellar medium (ISM).  In other words, the coherence of a modulated information-bearing signal would be limited in time or bandwidth extent by the combined effects of the relative motion of the source and receiver and the inhomogenous plasma occupying the intervening space.  Depending on the degree to which these effects can be corrected for, they define either limits or search parameters of any algorithm used to detect information bearing radio emission sent over interstellar distances. 
    
\begin{figure}[htb]
\begin{center}
\begin{tabular}{c}
\includegraphics[width=0.85\linewidth]{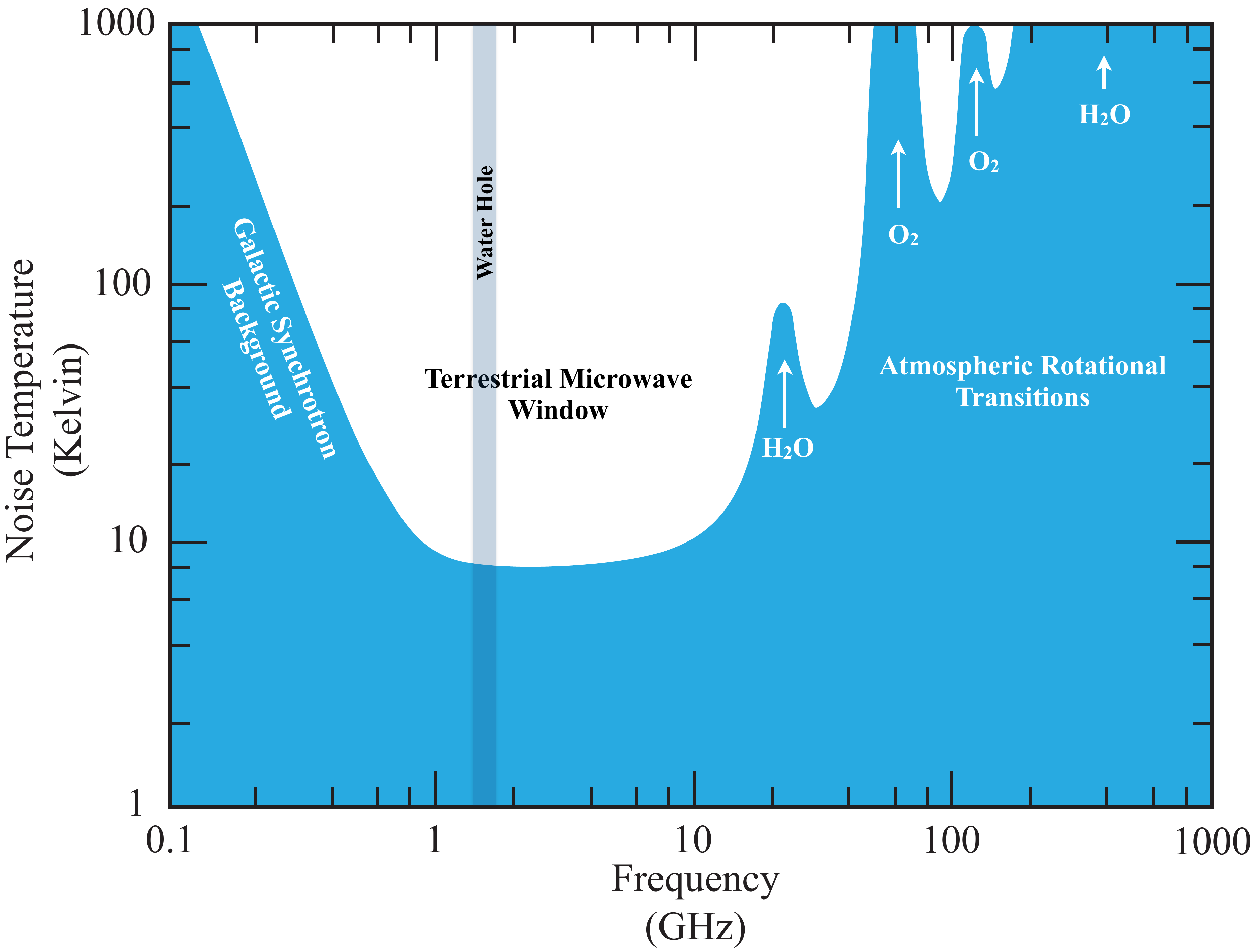}
\end{tabular}
\caption{\footnotesize{The ``terrestrial microwave window'': a relatively quiet radio window between non-thermal galactic synchrotron emission and molecular rotational transitions in the Earth's atmosphere.  Also indicated on the figure is the so-called ``Water Hole'' bound by the frequencies of hyperfine transitions of neutral hydrogen (H, $f_{\rm H} \approx$ 1.42 GHz) and the hydroxyl radical (OH, $f_{\rm OH} \approx$ 1.67 GHz) }}
\label{fig:window}
\end{center}  
\end{figure}

In addition to changing ideas about signal types, there has been steady growth in the number and diversity of suggestions of preferred frequencies for radio SETI.  The so-called ``terrestrial microwave window'' (TMW, Figure \ref{fig:window}), the spectral region of relatively low natural noise between the galactic synchrotron background and emission and absorption by water and oxygen in the Earth's atmosphere, was identified early as an ideal band to conduct radio SETI \citep{Morrison:1977p182}.  Faced with technological limitations that made it very difficult to search the entire $\sim$1$-$15 GHz window, early SETI experiments developed the notion of the ``water-hole,'' an especially attractive region of spectrum at the bottom of the TMW trough bound by the frequencies of hyperfine transitions of neutral hydrogen (H, $f_{\rm H} \approx$ 1.42 GHz) and the hydroxyl radical (OH, $f_{\rm OH} \approx$ 1.67 GHz).  In addition to the romantic allusion to terrestrial desert oases where life in a barren land gathers to sate, there was a convenient supply of large telescopes and sensitive radio receivers operating in that range already.  As time progressed, other ``special frequencies'' were identified.  \cite{1993A&A...278..669B} lists 55 individual ``interstellar communication channel'' frequencies between 500 MHz and 25 GHz based on scaling\footnote{e.g. $f_{\rm H}\pi=$4.462336273 GHz as given in the reference} the rest frequencies of H and OH hyperfine transitions by mathematically special ``civilization signature constants,'' e.g. $\pi, \pi/2, 2\pi, e, \pi e, e^e$.  \cite{1993ASPC...47..161G} identified additional preferred frequencies based on hyperfine transitions of neutral hydrogen in the $n=2$ excited state.  

Bands above and below the TMW present compelling opportunities as well.  \cite{2007JCAP...01..020L} have detailed a number of reasons why the band between 50$-$300 MHz might be worthwhile, including the fact that some of the Earth's most detectable emission is leaked at these frequencies.  While observations that encroach on the upper end of the TMW tend to become more difficult and less sensitive due to increased atmospheric noise, radio transmissions at these frequencies are less affected by the interstellar and interplanetary plasma and might be more easily produced at very high luminosities \citep{Benford:2010p615,2010AsBio..10..475B,2010AsBio..10..491B}.  \cite{1973R&QE...16.1118G} pointed out that the received signal to noise ratio (SNR) for pulse communication with delay compensation increases by more than an order of magnitude between 1 and 10 GHz, reaching a peak around 56 GHz.  \cite{1979Natur.278...28K} argued that advanced civilizations transmitting deliberate beacons might generally prefer frequencies higher than a few GHz due to less impairment from the ISM, and identified several particular ``magic frequencies'' related to the cosmic microwave background temperature and hyperfine transitions in the positronium ``atom'' around 200 GHz.  Fortunately modern radio astronomy technology allows searches over many hundreds of MHz to GHz of instantaneous bandwidth, lessening the extent to which these ``magic frequencies'' need be considered. 

As our own radio technology advances, we will undoubtedly develop new ideas about what type of emission we might expect from similar technology at work on another world.  Though there are certain frequencies and signal types that have been particularly advocated to-date, an ideal terrestrial radio SETI observatory would provide near continuous frequency coverage over a significant fractional bandwidth, and flexible signal detection systems that could be programmed to use multiple search algorithms.  Such a system is precisely what the SKA will provide - through an open and extensible digital signal processing backend, the SKA will enable SETI observers to readily program flexible compute elements with their own algorithms, and optionally attach user-supplied custom hardware if necessary.  The world's most sensitive radio telescope will thus be perfectly suited to conducting novel SETI observations, and as we detail below, will permit the most sensitive search for artificial radio emission ever conducted.

\section{SETI Observations on the SKA}
  
\subsection{Commensality}

Since the inception of modern radio SETI, observers have grappled with a fundamental problem.  The SETI search space is so broad in many dimensions, crucially including both frequency and sky location, that a proper search requires very large amounts of telescope time.  However, the dedication of significant amounts of observing time on the largest publicly-funded radio telescopes to a single science program, let alone one as speculative as SETI, is simply not possible.  More than 30 years ago, SETI astronomers devised a solution that they dubbed ``parasitic SETI'' \citep{Bowyer:1983p1498}.  By taking advantage of the fact that the amplified sky signal from a radio telescope can be duplicated many times over with very little added noise or loss of sensitivity, SETI astronomers could ``piggy-back'' on other users' observations to conduct vast sky surveys for signs of intelligent life without ever billing a minute of primary-user time.  Now given the friendly and more accurate name ``commensal observing,'' this technique is enabling the Search for Extraterrestrial Radio Emission from Nearby Developed Intelligent Populations (SERENDIP), SETI@home and AstroPulse projects to use the Arecibo Observatory to conduct SETI observations for thousands of hours per year.  Astronomers of all stripes have now begun to recognize the efficacy of commensal observing for a wide range of science programs that can take advantage of very large amounts of observing time without rigorous constraints on integration period or the need to target specific objects or fields.  Commensal searches for radio transients are now routinely conducted with the VLBA \citep{2011ApJ...735...97W}, and other commensal transient and SETI pipelines are in development or early operation at the JVLA, GBT, LOFAR \citep{2013IAUS..291..492S}, Jodrell Bank and elsewhere.  

\begin{figure}[tb]
\begin{center}
\begin{tabular}{c}
\includegraphics[width=0.85\linewidth]{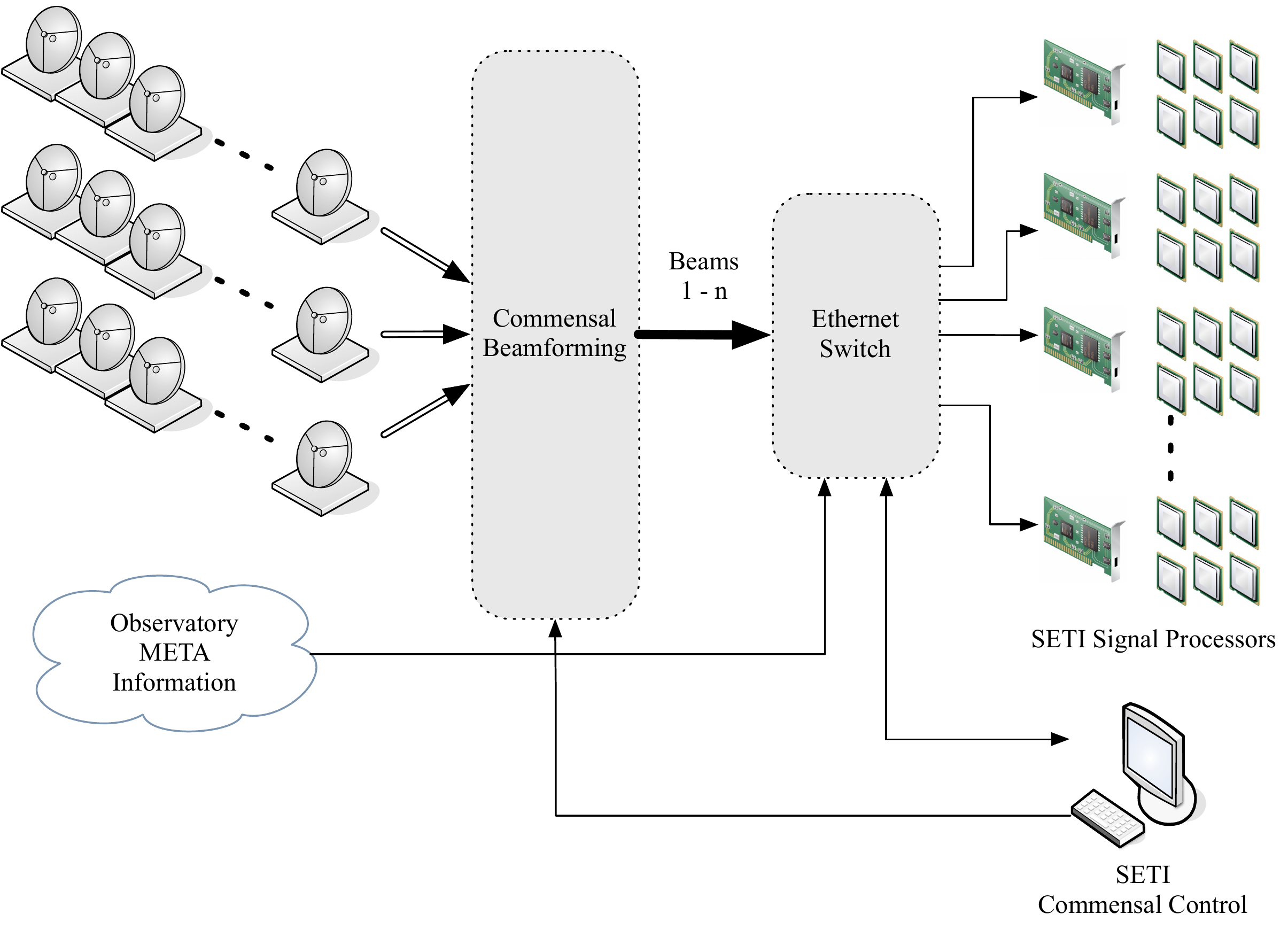}
\end{tabular}
\caption{\footnotesize{Schematic Diagram of Commensal Observation on the SKA: a SETI observer employs an independent beamforming capability to form phased-array beams within the current primary field(s) of view based on observatory meta information, and directs the unprocessed digital voltage data for those beams to SETI signal processor via Ethernet}}
\label{fig:commensal}
\end{center}  
\end{figure}

Support for commensal observations will be critical for the SKA SETI program.  For the discussion here, we define commensal observations to mean two or more users using the observatory largely independently, sharing only the primary field(s) of view chosen by the primary observers.  From a SETI perspective, we envision an observing system in which multiple phased array beams can be independently steered within the primary field-of-view and fed to dedicated signal processing hardware where SETI specific algorithms are implemented.  The Allen Telescope Array's designed mode of operation is an excellent schematic prototype for the way in which such a system can function \citep{2006AcAau..59.1153D, 2009IEEEP..97.1438W}.  Figure \ref{fig:commensal} illustrates this scheme: a SETI observer employs an independent beamforming capability to form phased-array beams within the current primary field(s) of view based on observatory meta information, and directs the unprocessed digital voltage data for those beams to SETI signal processors that perform interference excision and signal detection.  Using a technology such as Ethernet, multiple copies of a single beam can be directed to multiple signal processors for analysis using multiple techniques.  The most sensitive SETI observations with the SKA will be conducted in this phased-array mode, but it is worth pointing out that there are additional opportunities for conducting parallel SETI analysis on imaging and raw u-v data products as well.    

\subsection{Signal Processing}
\label{sec:sig}
Signal detection in current SETI experiments is essentially an exercise in matched filtering, in which the filter template is defined by the assumed characteristics of the transmitter.  As discussed in Section \ref{sec:are}, most radio SETI experiments focus on narrow-band emission, and thus the appropriate matched filter is a high resolution Fourier transform, combined with an additional step to account for the unknown relative acceleration between transmitter and receiver.  A suite of other algorithms might be effectively employed in a SETI experiment, including principal component analysis \citep{Biraud:1983gr, Maccone:1991ue}, autocorrelation \citep{Gardner:1992wg}, symbol-wise autocorrelation (SWAC)  \citep{Morrison:2011wf} and automatic modulation classification (AMC) \citep{Aslam:2010wo} and references therein, but are not yet in widespread use largely due to their computational cost.  SETI experiments on the SKA will benefit from significant increases in computational performance thanks to the continuing Moore's law growth in the electronics industry \citep{Moore:1965uj}, and thus will be able to consider implementing some of these techniques using the SKA's flexible signal detection systems.  Moreover, for the most part these techniques are ``embarrassingly parallel'' and can readily be scaled to accommodate enhanced capabilities.   

Here we will use the case of narrow-band detection as an illustrative example of SETI signal processing and for calculating the sensitivity of SKA SETI observations.  Although the sensitivity to sources of a given luminosity would vary somewhat for algorithms detecting other types of signals, the sensitivity expressions and calculations used here are roughly correct for other incoherent or weakly-coherent detection schemes \citep{SullivanIII:1984io, Gulkis:2018p1555}.

The minimum incoherently detectable flux, $F_i$, of a narrow band signal in one polarization is given by:
\begin{equation}
F_i  = S_i \Delta b_i = \sigma _{{\rm{thresh}}} S_{\rm{sys}} \sqrt {\frac{\Delta b}{t}} 
\end{equation}

Where $S_i$ is the intrinsic flux density of the transmitter, $\Delta b_i$ the intrinsic bandwidth of the transmitter, $\sigma _{{\rm{thresh}}}$ is the SNR,  $S_{\rm{sys}}$ is the system equivalent flux density (SEFD) of the receiving telescope, $\Delta b$ is the spectral channel bandwidth and $t$ the integration time.  Here we assume $\Delta b_i < \Delta b$, and have expressed $F_i$ in terms of flux rather than flux density to more intuitively relate experimental sensitivity to terrestrial transmitter power levels, which are commonly described by total apparent luminosity (or EIRP).    

For narrowband signals, the ``Doppler drift,'' or changing apparent frequency due to the relative acceleration between the transmitter and receiver is given simply by: 
\begin{equation}
\dot f = \frac{{d\overrightarrow V }}{{dt}}\frac{{f_{\rm rest} }}{c}
\end{equation}
where $\overrightarrow V$ is the line of sight relative velocity between receiver and source, $f_{apparent}$ is the apparent frequency of the transmitter for constant velocity and $c$ the speed of light.  Because we are searching for emission at an unknown frequency, the overall Doppler shift imposed on a signal due to some unknown constant relative velocity is unimportant for detection.  As points of reference, the maximum contribution from the Earth's orbital motion at 1 GHz is $\sim  \pm 0.02 \rm{\ Hz\ s}^{-1}$, and from the Earth's rotation is\footnote{The maximum rotational contribution is negative because it occurs for a source observed from the equator at zenith.} $\sim  - 0.1 \rm{\ Hz\ s}^{-1}$. 

Narrow band signals are spectrally broadened by both the ISM and interplanetary medium (IPM), with a magnitude equal to \citep{1991ApJ...376..123C, 2013ApJ...767...94S}: 
\begin{equation}
\label{eq:dnu}
\Delta \nu _{{\rm{broad-ISM}}}  = 0.097{\rm{\ Hz\ }}\nu _{{\rm{GHz}}} ^{ - 6/5} {\left( {\frac{{V_ \bot  }}{{100}}} \right)} {\rm{SM}}^{ 3/5} 
\end{equation}
and
\begin{equation}
\label{eq:dnusolar}
\Delta \nu _{{\rm{broad-IPM}}}  = 300 {\rm{\ Hz\ }}\nu _{{\rm{GHz}}} ^{ - 6/5} {\left( {\frac{R}{R_{\odot}}} \right)} ^{-9/5}
\end{equation}

Where $\nu _{{\rm{GHz}}}$ is the observing frequency in GHz, $R$ is the Solar impact distance, $V_{\bot}$ is the transverse (perpendicular to the line of sight) velocity of the source in km/sec and SM the direction dependent scattering measure, a measure of the electron density fluctuations (c.f. \citealp{1990ARA&A..28..561R}) integrated along the line of sight.  For most lines of sight with $R>\sim100\ R_{\odot}$ out to several kpc, the total spectral broadening is no more than\footnote{$V_{\bot}$ = 25 km/sec, SM from \cite{2002astro.ph..7156C}} $\sim$ 0.1 Hz$\ \nu _{{\rm{GHz}}} ^{ - 6/5} $  

The combined influence of the above effects thus dictate the parameters of an optimal search for narrow-band transmitters.  Around 1 GHz for example, frequency resolutions of order 0.1 Hz and (unknown) drift correction up to a few Hz/sec would permit optimum detection of equatorial transmitters on a planet up to five times larger and rotating five times faster than the Earth at distances of around 1 kpc.  Algorithms used to detect other types of signals will be similarly parameterized, in that they will search or average over similar parameters and will also be constrained based on the expected properties of the signal and communication channel.              

\subsection{Target Selection}
Generally speaking there are two possible SETI observing strategies, targeted observations of individual objects or fields thought to be especially likely to host intelligent life or blind surveys of large swaths of the sky.  Occasionally hybrid approaches are employed, e.g. surveys of the {\it Kepler} Field \citep{2013ApJ...767...94S} or the Galactic Center region \citep{1985AcAau..12..369S}.  The relative merit of these two approaches depends strongly on the luminosity function of artificial radio transmitters, which at present is largely unconstrained \citep{Tarter:2004p258}, although some insight can be drawn from examining the luminosity distribution of terrestrial transmitters (See Section \ref{sec:sens}).  This fact, combined with many other unknowns, will lead to a wide range of views regarding how SETI observations should be conducted on the SKA.  As is the case with any science program, these proposals should be subject to peer review and those targets deemed most compelling will be observed with highest priority.  \cite{Tarter:2001hi} and references therein detail the varied proposed approaches to SETI target selection in detail.  

In practice, the vast majority of SETI observing on the SKA will be conducted commensally with other science programs, employing a signal processing system that will be capable of conducting searches using multiple phased-array beams over only a fraction of the primary beam.  Thus an all-sky catalog providing a number of targets per primary field of view, at the highest frequencies, at least equal to the number of phased-array SETI beams (2$-$4 in the minimal case), is required.  Such a catalog would consist of $\sim$ 25M targets, and could be readily constructed from the expected results of {\it GAIA} \citep{2001A&A...369..339P} following a prescription similar to that used in the ``HabCat'' catalogue of habitable stars \citep{Turnbull:2003p273,2003ApJS..149..423T}.  That is, essentially choosing stars and exoplanet systems that appear most conducive to life as we know it, or in other words are similar to our own Solar system.  

However, a HabCat-like catalog should be supplemented using additional strategies.   For example, a simple isotropic volume-selected sample of stars is an ideal target set for a SETI experiment for a number of reasons.  Most importantly, it ameliorates the inherent anthropocentric bias that can lead us to concentrate SETI efforts on environments similar to our own planetary system.  A search of a blind volume-limited sample of stars is premised only on the notion that intelligent life requires a stellar host in order to develop.  It also probes a diverse nearby stellar population that is exceedingly well studied at many wavelengths.  In the event of a non detection, these attributes of the sample will allow us to place strong and broadly applicable limits on the presence of technologically-produced radio emission.  It also allows us to search very deeply, which may be important if the artificial radio transmitter luminosity function declines steeply much past our own technologically produced radio luminosity.

\begin{figure}[htb]
\begin{center}
\begin{tabular}{c}
\includegraphics[width=0.7\linewidth]{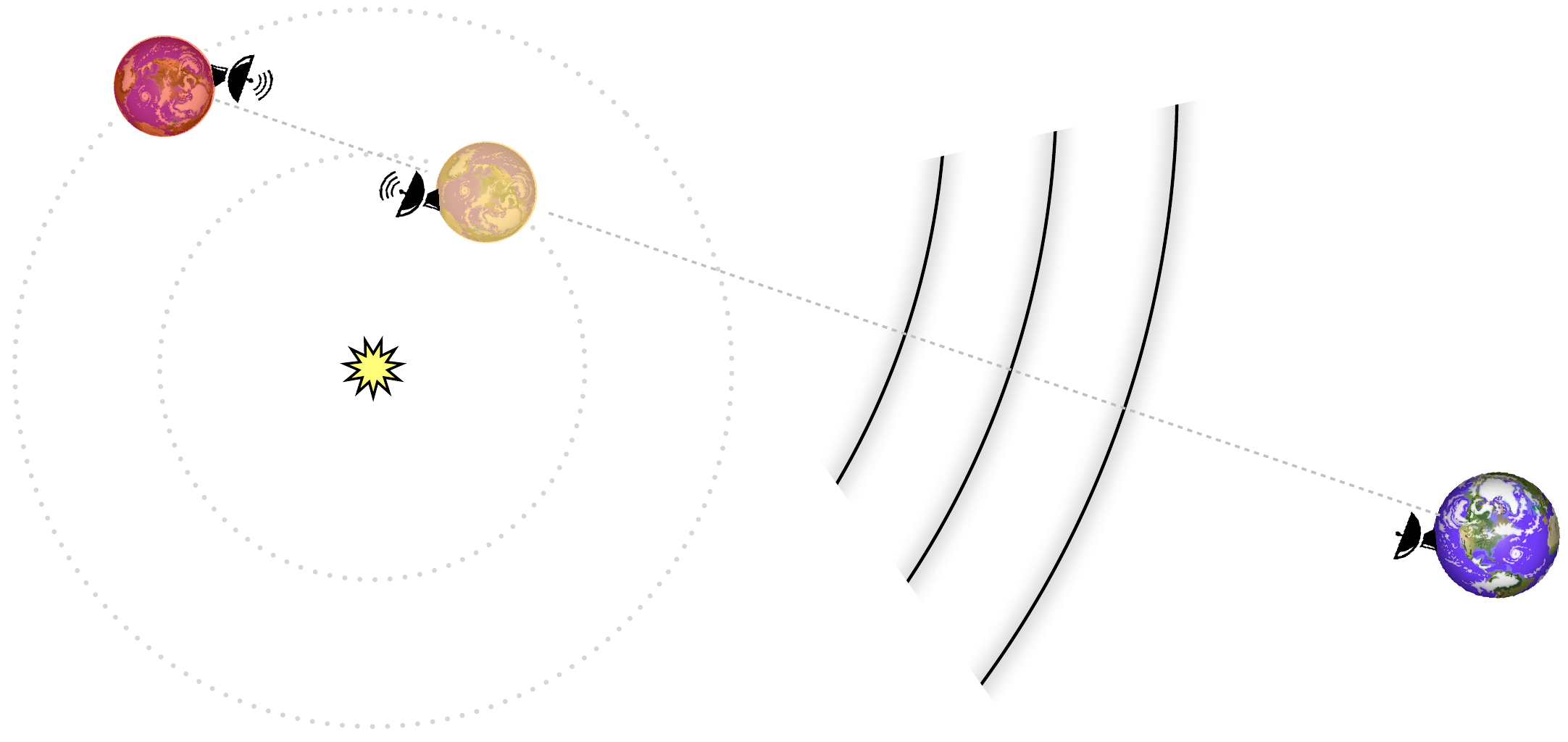}
\end{tabular}
\caption{\footnotesize{An illustration of a 2-planet conjunction in an extrasolar planetary system along a line of sight to the Earth. \vspace{-25px}}}
\label{fig:alignment}
\end{center}
\end{figure}

The plethora of multi-planet exoplanet systems now known present a unique opportunity for performing SETI searches at particularly advantageous times - epochs of conjunction along a line of sight to the Earth (See Figure \ref{fig:alignment}).  Extrapolating from humanity's exploration of space, it is likely that a more advanced civilization having similar proclivities would explore and perhaps colonize multiple planets in their star system.  These explorations could very easily include planet-planet communication, radar imaging or radar mapping of orbital debris, perhaps similar to the way in which we use the Arecibo Planetary Radar to image objects in our own Solar System.  Observing these systems during alignments would allow us to potentially eavesdrop on this emission.  The published ephemerides of these known systems readily enable calculation of conjunction times, and many more such systems are expected to be discovered in the coming decades.

\begin{figure}[htb]
\begin{center}
\begin{tabular}{c}
\includegraphics[width=0.9\linewidth]{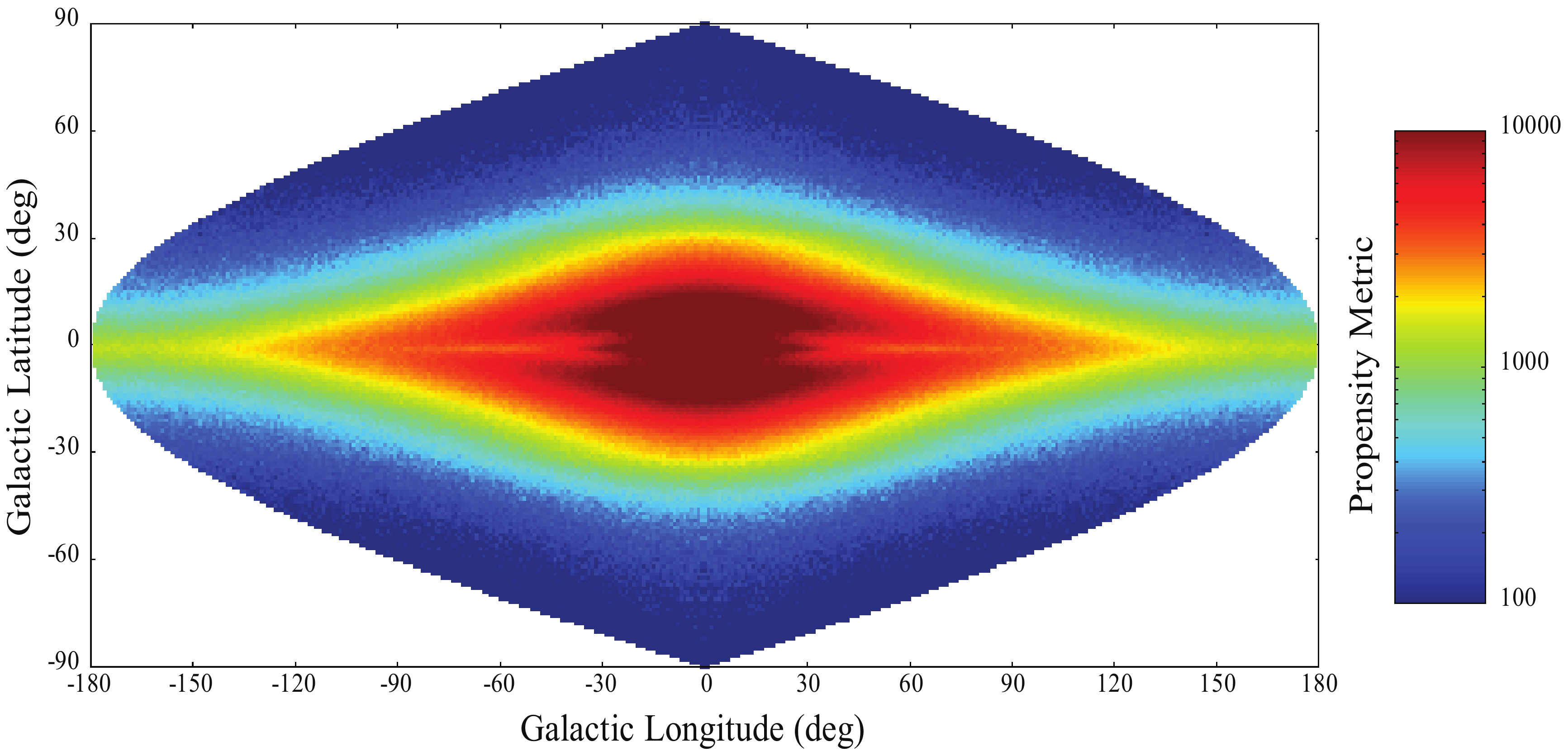}
\end{tabular}
\caption{\footnotesize{Contour map plot of the relative propensity for the emergence of intelligence as a function of telescope pointing direction in galactic coordinates, based on supernova event rates.  An equal-area sinusoidal projection is employed, where each pixel represents approximately one square degree on the sky. For each pixel, the plotted metric (arbitrary units) is proportional to the accumulated time available for intelligence to evolve across all habitable planets within that pixel out to a maximum range of 5 kpc from Earth, summed over all epochs. The available times for intelligence to emerge occur during gaps between nearby supernova events.  Only gap times exceeding a minimum threshold are included in the summations. For this plot, a threshold of 2.15 Gyr is employed, consisting of 1.55 Gyr (the assumed time for land-based complex life to evolve) plus 0.6 Gyr (the assumed minimum time for complex life to further evolve to intelligence).  \vspace{-25px}}}
\label{fig:ghz}
\end{center}
\end{figure}

Another idea that may inform the construction of an SKA SETI catalog is the notion that perhaps there are preferred regions of the galaxy for the development of complex life, so-called ``galactic habitable zones'' (GHZ), e.g. based on stellar population similarity to the Sun or a low rate and proximity of supernovae \citep{2004Sci...303...59L}.  Figure \ref{fig:ghz} depicts one possible realization of such a GHZ from \cite{S:2014ul}.  For each pixel shown, the plotted metric (arbitrary units) is proportional to the accumulated time available for intelligence to evolve across all habitable planets within that pixel out to a maximum range of 5 kpc from Earth, summed over all epochs. The available times for intelligence to putatively emerge occur during gaps between nearby supernova events.  As shown, this work suggests that a region of the sky centered on the galactic centre and spanning approximately 60$^{\circ}$ of longitude and 30$^{\circ}$ of latitude is especially attractive.  Other additions to a SKA SETI target list might include natural sources that could serve either as signposts or natural amplifiers, e.g. using a maser filament to amplify a narrow-band signal transmitted at a specific maser transition frequency \citep{1993ASPC...47..257C, 1995ASPC...74..325C}.    

\subsection{Sensitivity}
\label{sec:sens}
\begin{figure}[htb]
\begin{center}
\begin{tabular}{c}
\includegraphics[width=0.9\linewidth]{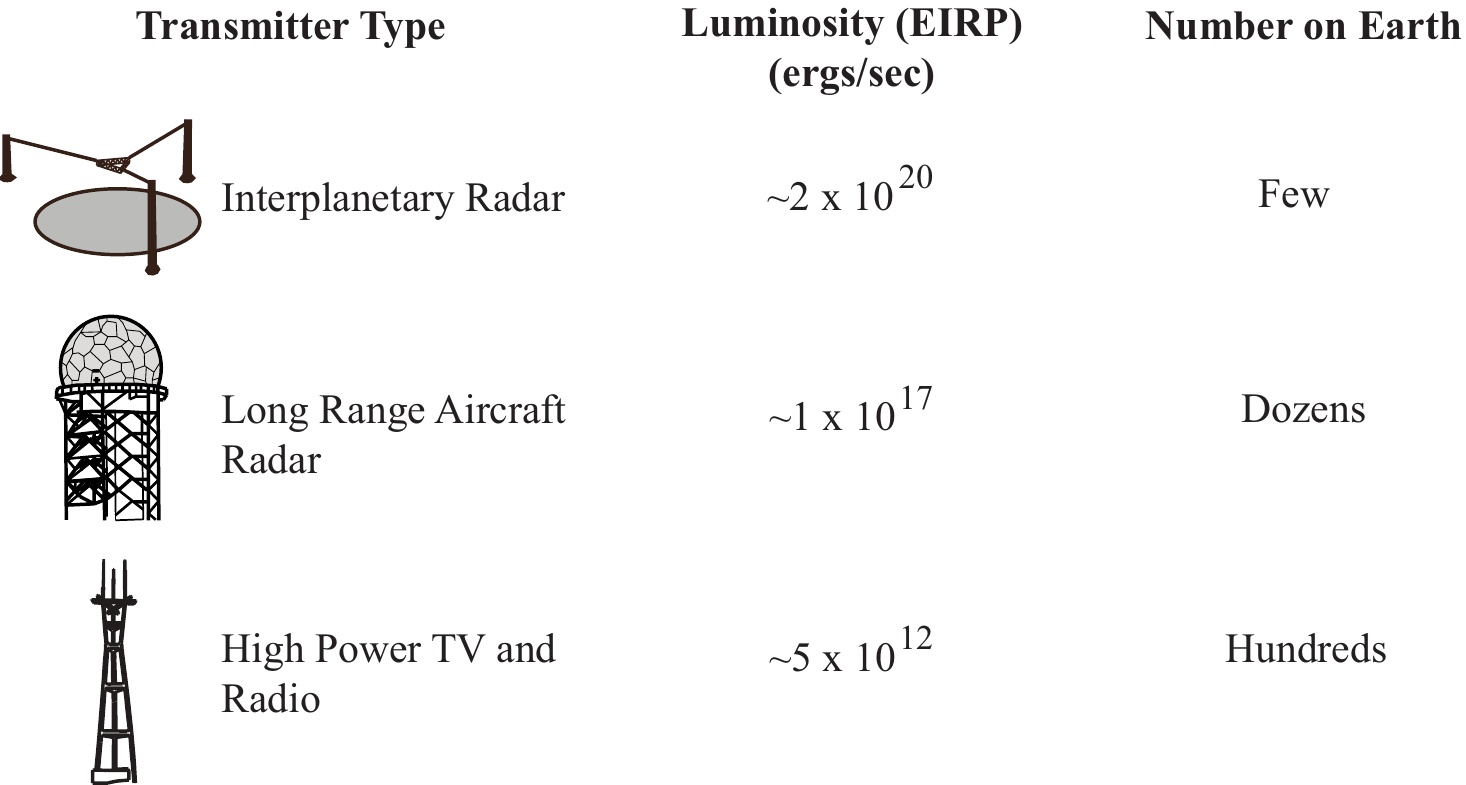}
\end{tabular}
\caption{\footnotesize{A table of terrestrial analogs of artificial extraterrestrial radio sources, including the pseudo-luminosity, as expressed by the equivalent isotropically radiated power (EIRP) and the approximate number of such transmitters present on Earth.\vspace{-25px}}}
\label{fig:terrestrial}
\end{center}
\end{figure}

Absent of the actual detection of an artificial extraterrestrial radio transmitter, our best points of reference for the sensitivity of radio SETI experiments, and the luminosities of sources we might detect, come from our own terrestrial technology.  Figure \ref{fig:terrestrial} lists several terrestrial transmitters that produce emission in the bands probed by the SKA, along with their pseudo-luminosities as described by their equivalent isotropically radiated power (EIRP).  The approximate number of such transmitters on the Earth is also listed.  These terrestrial analogs of artificial transmitters are included simply to describe the energetics of transmitters that SKA SETI experiments might be sensitive to in an intuitive way.  Although in many cases these transmitters produce emission precisely alike that searched for in radio SETI experiments, in other cases they modulate their output power in such a way that they would not be as detectable as is implied by considering their transmissions as a simple sinusoid.

\begin{figure}[htb]
\begin{center}
\begin{tabular}{c}
\includegraphics[width=0.99\linewidth]{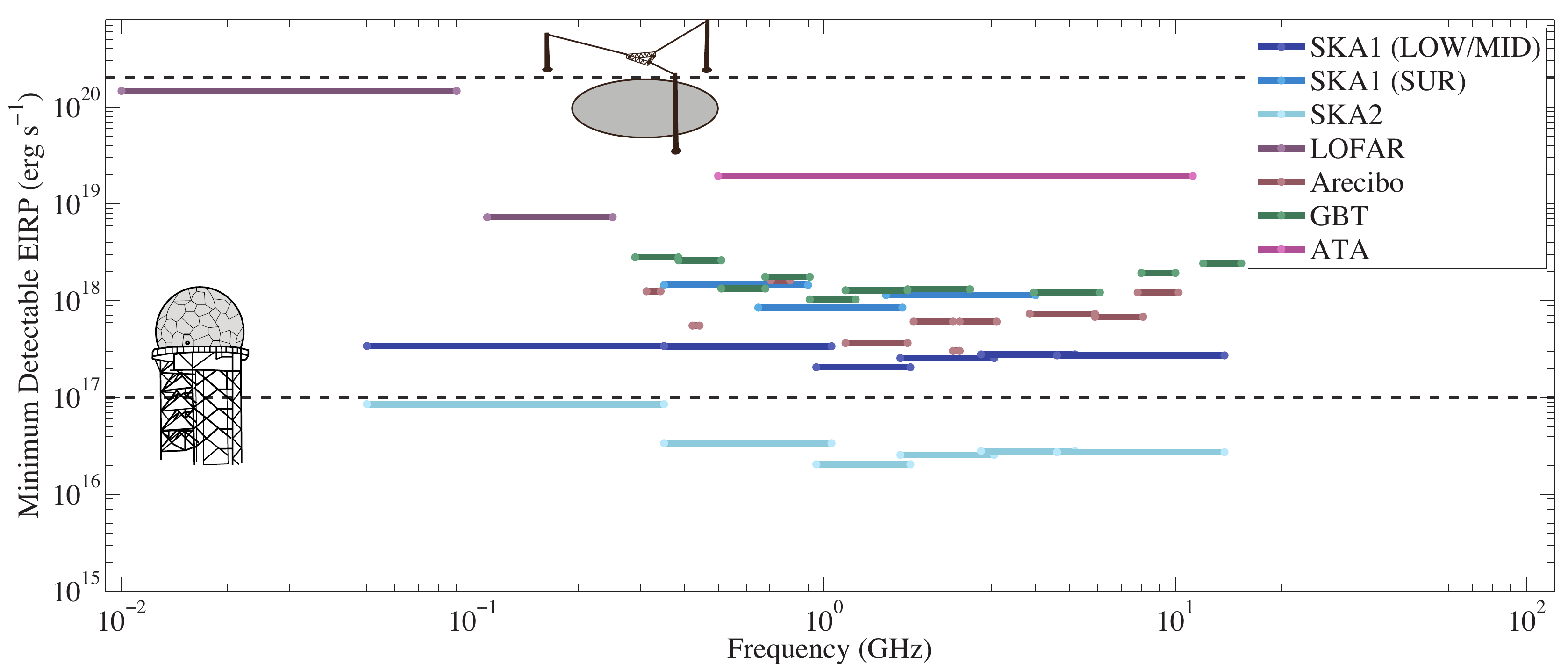}
\end{tabular}
\caption{\footnotesize{Sensitivity of each component of the SKA to narrow-band transmitters at 15 pc, as compared with other facilities actively performing SETI searches over the same band.  Here we assume a significance threshold $\sigma=15$, bandwidth $\Delta b=0.5$ Hz and integration time $t=10$ minutes.   A transmitter is detectable if its EIRP is above the curve for a given telescope.  Thus a transmitter with an EIRP of 2 x 10$^{20}$ ergs/sec (planetary radar) is detectable with all of the telescopes shown, while a transmitter with an EIRP of 1 x 10$^{17}$ ergs/sec (airport radar) is detectable only with SKA2.  Sensitivities for the Green Bank Telescope (GBT), Arecibo and LOFAR were taken from those facilities' observing guides.  For LOFAR we assume only core stations are used.  Allen Telescope Array (ATA) sensitivity was taken from \cite{2009pra..confE...5V}. }}
\label{fig:sensitivity}
\end{center}
\end{figure}

Figure \ref{fig:sensitivity} depicts the sensitivity of each component of the SKA to narrow-band transmitters at 15 pc, as compared with other facilities actively performing SETI searches over the same band.  Search parameter assumptions here match roughly what might be expected for a significant fraction of commensal observations, namely a maximum integration time of 10 minutes.  As shown, with appropriate signal detection systems the raw sensitivity of the SKA can be leveraged to create the most sensitive SETI system in the world.  In Figure \ref{fig:sensitivity}, a transmitter is detectable if its EIRP is above the curve for a given telescope.  Thus in the observing scenario presented, a transmitter with an EIRP of 2 x 10$^{20}$ ergs/sec (planetary radar) is detectable with all of the telescopes shown, while a transmitter with an EIRP of 1 x 10$^{17}$ ergs/sec (airport radar) is detectable only with SKA2.  

\begin{figure}[htb]
\begin{center}
\begin{tabular}{c}
\includegraphics[width=0.99\linewidth]{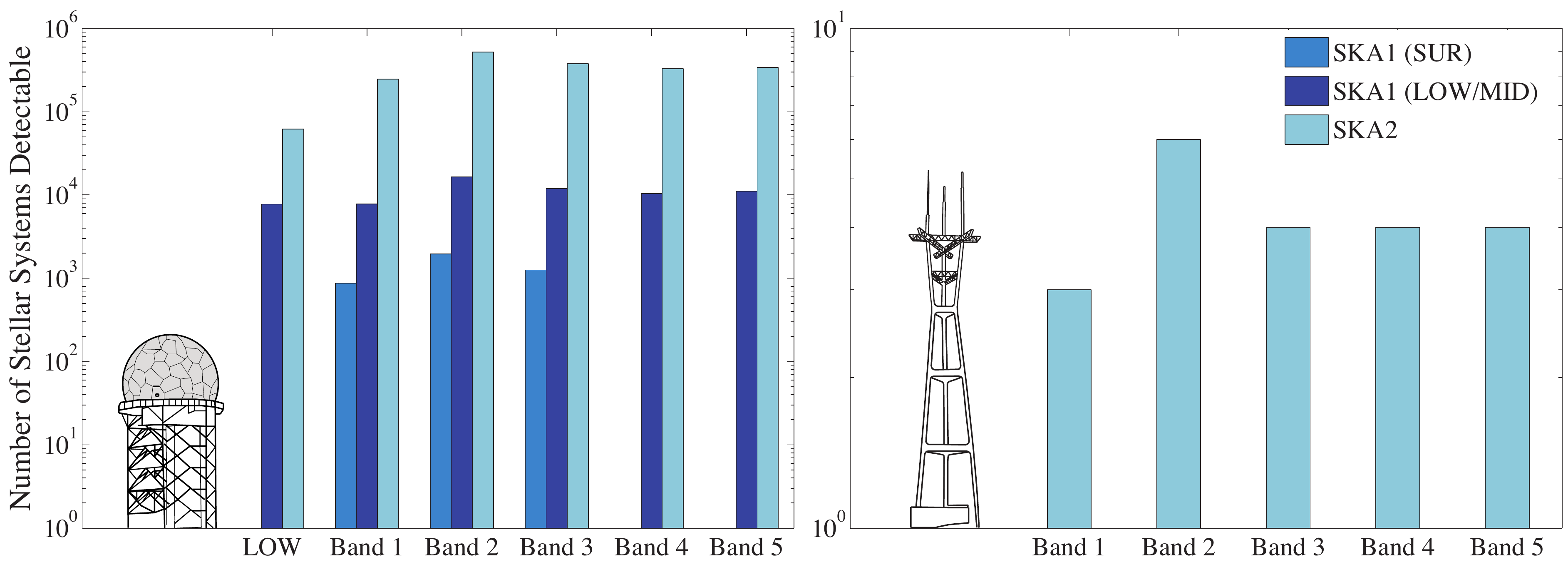}
\end{tabular}
\caption{\footnotesize{The number of stars in the solar neighborhood from which narrow-band emission would be detectable at two luminosity levels.  Here we assume a stellar number density of $n_{*}$ = 0.1 pc$^{-3}$, significance threshold $\sigma=12$ and integration time $t=60$ minutes.  For luminosities similar to our terrestrial aircraft radars we assume bandwidth $\Delta b=0.01$ Hz and for luminosities similar to terrestrial TV/radio-like signals (right) we assume fully coherent integration ($\Delta b=t^{-1}$).   The search parameters shown here, namely the smaller value of $\Delta b$ than discussed in Section 2.2, reflect the relative proximity of these nearby stars and by extension their less scattered lines of sight.}}
\label{fig:numstars}
\end{center}  
\end{figure}

Figure \ref{fig:numstars} depicts what sensitivities could be attained in a more optimistic scenario, in which SETI was the primary observing purpose or commensal observations were performed with another science case very well matched to SETI.  Here we assume an integration time of 60 minutes, and the minimum channelization bandwidth permitted by ISM and IPM effects.  As shown, with SKA1, radio transmitter luminosities similar to our high power radars will be detectable from tens of thousands of stars across the entire terrestrial microwave window, and with SKA2 these signals will be detectable from hundreds of thousands of stars.  Further, with SKA2 we will for the first time have the sensitivity to detect radio emission similar in power to our own TV and radio stations from a few of our nearest neighbors.

As is the case with many large scientific facilities, we expect the full capabilities of the SKA to emerge over time.  SETI is very well suited to conducting early deployment science, as the breadth of unexplored parameter space in SETI is quite large in terms of frequency coverage and raw sensitivity.  Although increased sensitivity and digital capabilities, especially the availability of additional phased-array beams, will increase the depth and breadth of SETI on the SKA, with 50\% build-out of SKA1 across the board a SETI program could begin to probe the nearer targets in a luminosity-limited survey.  More distant targets could be reserved for full sensitivity.  However, it is worth pointing out, as shown in Figure \ref{fig:numstars}, that only SKA2 has the ability to detect TV and radio station-like transmitters.  With the expected number of planetary systems for which it would be sensitive to these in the single digits, any reduction in sensitivity of SKA2 could render a nearby human-like civilization invisible.

\section{Summary}
SETI searches with the SKA1 facilities will build from an active base of theoretical and experimental work being done in the field with existing large-scale facilities, but the combination of raw sensitivity, flexible electronics and increased computational capacity will enable orders-of-magnitude improvement in the speed, depth and breadth of previous SETI experiments.  The most thorough targeted SETI search previously conducted surveyed 1000 stars over 1$-$3 GHz to a luminosity limit of $\sim$2 x $10^{19}$ ergs/sec \cite{1998AcAau..42..651B}.  In a five year commensal campaign, SKA1 could survey every star in its declination range within 60 pc $-$ more than ten thousand stars $-$ to a luminosity limit an order of magnitude fainter, $\sim10^{18}$ ergs/sec, over a larger band. Conservatively, in ten years SKA2 could survey every star within 60 pc to a luminosity limit equal to the EIRP of terrestrial aircraft radars over the entire terrestrial microwave window.
  
Historically, performing SETI experiments effectively with large telescopes has been technically and politically difficult.  SKA will be the first world-class telescope ever constructed for which SETI observations are a conscious part of the design process and are available as a facility observing mode.  Our most powerful tool to search for other intelligent life in the cosmos will be available and accessible to astronomers around the world, infusing new people and ideas in to the field of SETI and maximizing the scientific return of the SKA.  Searches for life beyond Earth, especially searches for intelligent life, have a remarkable ability to spark the imagination of the public, and this component of the SKA science case will offer ample opportunity for outreach and public engagement.  From a purely scientific standpoint, the discovery of an independent genesis of life on another world would provide strong evidence that life is common, and its intelligence would suggest that evolution proceeds towards this end easily.  Were an information-containing transmission to be received and decoded, we can only conjecture what thoughts and ideas might be contained therein, but surely our human culture would be enriched in unimaginable ways.\\

\vspace{5px}\noindent{\bf Acknowledgements  }
We thank an anonymous referee for thoughtful comments on this work.  Part of this research was carried out at the Jet Propulsion Laboratory, California Institute of Technology, under a contract with the National Aeronautics and Space Administration.  A. P. V. S., E. J. K. and D. W. received partial support from a competitive grant awarded by the John Templeton Foundation.

\bibliographystyle{apj}

\bibliography{references}

\end{document}